\documentclass[12pt,preprint]{aastex}

\slugcomment{}

\shorttitle{AlO molecular bands in  V4332 Sgr}
\shortauthors{AAAA et al.}

\begin{document}


\title{Remarkable changes in the near-infrared spectrum of the 
nova-like variable V4332 Sgr}


\author{Dipankar P.K. Banerjee}
\affil{Physical Research Laboratory, Navrangpura,  Ahmedabad, India, 380009}
\email{orion@prl.ernet.in}
\author{Watson P. Varricatt}
\affil{Joint Astronomy Center, 660 N. Aohoku Place, Hilo, Hawaii-96720, USA}
\email{w.varricatt@jach.hawaii.edu}
\author{Nagarhalli M. Ashok}
\affil{Physical Research Laboratory, Navrangpura,  Ahmedabad, India, 380009}
\email{ashok@prl.ernet.in}
\and

\author{Olli Launila}
\affil{SCFAB-KTH, Atomic and Molecular Physics, Roslagstullsbacken 21,
 SE-106 91, Stockholm, Sweden}
\email{olli@physics.kth.se}


\begin{abstract}
        We report on recent near-IR observations of V4332 Sgr - the
        nova-like variable that erupted in 1994. Its rapid, post-outburst
        evolution to a cool M type giant/supergiant, soon after its 
        outburst,      had showed that it was an
        unusual object differing from other eruptive variables like
        classical/symbiotic novae or born-again AGB stars.  The present
        study of V4332 Sgr was motivated by the keen interest in the
        recent eruption of V838 Mon - an object with a spectacular
        light-echo and which, along with V4332 Sgr, is believed to belong
        to a new class of objects (we propose they may be called 
        ``quasi-novae''). Our observations show new developments
        in the evolution of V4332 Sgr. The most striking feature is the
        detection of several molecular bands of AlO  - a rarely seen
        molecule in astronomical spectra - in the $JHK$ spectra. Many of
        these bands are being detected for the first time.  The only other
        detection of some of these AlO bands are in V838 Mon, thereby
        showing further spectral similarities between the two objects.
        $JHK$ photometry shows the development of a new dust shell around
        V4332 Sgr with a temperature of $\sim$ 900K and a lower limit on
        the derived mass of
        $M{_{\rm dust}} = 3.7 \times 10{^{\rm -12}}$$M${$_\odot$}. 
        This dust shell does not appear to be associated with ejecta
        of the 1994 outburst but is due to a second mass-loss episode
        which is not expected in a classical nova outburst. The cold
        molecular environment, suggested by the AlO emission, is also not
        expected in novae ejecta. We model the AlO bands and also discuss
        the possible formation mechanism of the AlO. These results show
        the need to monitor V4332 Sgr regularly - for unexpected
        developments. The results can also be significant in suggesting
        possible changes in the future evolution of V838 Mon.
	
\end{abstract}




\section{Introduction}
        V4332 Sagittarii (V4332 Sgr) erupted in a nova-like outburst in
        February, 1994. The light curve of the object (Martini et al., 1999)
        showed a slow rise in  brightness from an estimated, pre-outburst 
        magnitude of $B$ $\sim$ 18 to an extended period of  maximum brightness
        with $V$ = 8.5. The only detailed study of the object by
         Martini et al. (1999)  showed that its outburst does
        not conform to known categories of eruptive variables. V4332 Sgr
        showed a rapid cooling over three months - from 4400 to 2300K -
        and evolved into a cool M giant/supergiant.  A similar behaviour has
        also been observed in the recent (January 2002) outburst of V838
        Monocerotis. It appears, that V838 Mon and V4332 Sgr are analogues
        and together may be defining a new class of objects (Munari et al.
        2002; Bond et al. 2003). A third, possible member of this class is
        a red variable that erupted in M31 called M31 RV (Rich et al.
        1989).  While the outburst mechanism of such objects is not well
        understood, a recent result by Soker and Tylenda (2003) explains
        the outburst as a merger event between two stars - the release of
        gravitational energy leading to the the nova-like outburst. This
        is in distinct contrast to the classical nova scenario wherein a
        thermo-nuclear runaway on the surface of the white dwarf,
        accreting matter from its secondary companion, leads to the nova
        eruption.  The limited studies on V4332 Sgr (with none in the IR)
        - and also the current interest in V838 Mon - prompted us to make
        the present observations of V4332 Sgr to see how it had evolved
        with time.  These observations, which give interesting results, are
        reported here.

\section{Observations}
        Spectroscopic observations were done with the 3.8m United Kingdom
        IR Telescope (UKIRT), Hawaii and the 1-5 micron Imager
        Spectrometer (UIST) using a 1024$\times$1024 InSb array.
        Observations were done by nodding the telescope on two positions
        separated by 12$\arcsec$ along a 4 pixel (0.48$\arcsec$) wide
        slit.  Flat fielding was done with a black body mounted inside the
        instrument and spectral calibration was done using an Argon lamp.
        Spectra of the source in the $J$, $H$ and $K$ bands were obtained
        on 2003 April 24. A second $J$ band spectrum was obtained on 2003
        May 4.  Table 1 gives the details of the spectroscopic
        observations. Because of possible slit-losses, we have calibrated
        the spectra using the photometry of 2003 June 19 assuming
        constancy in source brightness over the 56 day gap between
        the spectroscopic and photometric observations. The UKIRT
        photometric observations were done using UFTI (UKIRT Fast-Track
        Imager) which uses a 1024$\times$1024 HgCdTe array and gives a
        plate scale of 0.091$\arcsec$/pixel. The sky was photometric with
        a $K$ band seeing of 0.69$\arcsec$.  Observations in the $JHK$
        bands were done by dithering the object in several positions.
        Separate dark observations were done and flat field corrections
        were applied using flat fields generated from the object
        observations by median combining the observed frames.  The UKIRT
        standard star FS35 was observed under similar conditions and used for
        flux calibration. The details of the observations and the derived
        $JHK$ magnitudes are given in Table 2.

	\section{Results}
	\subsection{$JHK$ spectroscopy}
	
         We show in Figure 1 the $JHK$ spectra of V4332 Sgr. The most
        striking feature of the spectra is the presence of several,
        strong, molecular bands in emission. Based on laboratory spectra
        by Launila $\&$ Jonnson (1994) we have identified the bands to be
        due to rotational-vibrational transitions in the $A$$^{\rm
        2}$$\Pi$$_{i}$$-$$X$$^{\rm 2}$$\Sigma$$^{\rm +}$ band system of
        AlO. The $A$$^{\rm 2}$$\Pi$$_{\rm i}$ state of AlO is a low lying
        electronic state - the $A$$^{\rm 2}$$\Pi$$_{\rm 3/2}$, $A$$^{\rm
        2}$$\Pi$$_{\rm 1/2}$ states being typically 1 ev above the ground
        level. To facilitate a comparison between the observed and
        predicted bands, a synthetic spectrum of AlO (discussed later in
        more details) is also shown in Figure 1. We have listed in Table 3
        the laboratory wavelengths of the different bands whose detection
        seems to be certain and also marked their positions in Figure 1. 
        The good match between the observed and expected wavelengths
        makes the identification of the AlO bands secure.  At optical
        wavelengths the detection of AlO in astronomical objects is rather
        rare - a  notable exception being its detection in the cool, 
        circumstellar environment of the peculiar red giant star U Equulei 
       ( Barnbaum et al., 1996). In the near-IR, the only detection of
        AlO is very recent and is
        coincidentally seen in an object similar to V4332 Sgr, i.e. V838
        Mon (Geballe et al. 2002, Banerjee \& Ashok 2002b, Bernstein et
        al. 2003). There are no published results on the AlO bands in V838
        Mon as yet. Further, only some of the AlO bands reported here are
        seen in V838 Mon viz. - the (1,0) and (4,0) bands - the other
        bands reported here are new detections.

	\subsection{Spectral Energy Distribution}
       
               The spectral energy distribution (SED) of V4332 Sgr (Figure
        2) shows considerable change between 2MASS observations ($J$ =
        12.1 , $H$ = 11.6, $K$${_{\rm s}}$ = 10.992; epoch: 18 May 1998)
        and now.  The $JHK$ fluxes in Figure 2 have been computed after
        reddening corrections adopting $E(B-V)$= 0.32 from Martini et al.
        (1999) and using zero magnitude fluxes and the standard relations
        between $A$${_{\rm V}}$, $A$${_{\rm J}}$, $A$${_{\rm H}}$ and
        $A$${_{\rm K}}$ from Koornneef (1983). A 3250K blackbody curve
        fits the 2MASS data fairly well. The present data shows a steep
        rise towards the red. A black-body fit between 900-1000K gives the
        most reasonable fit to the data though there is some deviation in
        the $J$ band. It is possible that a hotter black-body component is
        also present, probably due to some flux from the central star seen
        in the 2MASS data. A multi-component black-body fit (with
        temperatures of 900K and 3250K) is also shown which does suggest
        the presence of a hot component. But to infer anything more on the
        hot component is difficult from the present data.  What is however
        clear, is the presence of a cool 900-1000K component, which we
        interpret to arise from a dust shell that has formed subsequent to
        the 2MASS observations. The luminosity of the dust shell
        $L$${_{\rm dust}}$ can be found by numerically integrating the
        900K SED curve and multipying by 4$\pi$$d$$^{\rm 2}$ - where $d$
        is the distance to V4332 Sgr.  Martini et al. (1999) have
        estimated an upper limit of $d$ = 300pc for V4332 Sgr. For such a
        value of $d$, $L$${_{\rm dust}}$ is found to be $\sim$
        0.85$L$$_\odot$ . Since the dust shell is reasonably well
        approximated by a black-body we equate $L$${_{\rm dust}}$ =
        4$\pi$$R$${_{\rm shell}}$$^{\rm 2}$$\sigma$$T$${_{\rm dust}}{^{\rm
        4}}$ and get a shell radius $R$${_{\rm shell}}$ of $\sim$
        35$R$$_\odot$ (where $\sigma$ is the Stefan-Boltzmann constant and
        $T$${_{\rm dust}}$ is the dust temperature).  This shell cannot
        be associated with the ejecta from the 1994 outburst which had an
        expansion velocity in the range 200-300 km/s (from the H$\alpha$
        line widths of Martini et al. (1999)) and, in the absence of any
        deceleration, should have a radius of $\sim$ $10$${^{\rm
        5}}$$R$$_\odot$ at present. This disagrees too severely with the
        derived shell radius. The distance $d$ has to be boosted to an
        unreasonably large value of $\sim$ 1 Mpc (making the object
        extra-galactic) to have consistency between the derived shell size
        from luminosity and kinematic arguments. It therefore appears that the
        matter in the shell is not associated with the 1994 outburst but
        involves a second episode of mass-loss. Some additional support for 
        this conclusion is that the angular diameter of the shell from the 1994 
        outburst, should be $\sim$  3.0$\arcsec$ (for $d$ = 300pc) at present
        and resolvable in the K band images (Table 2) which taken at
        0.69$\arcsec$ seeing - this is not found to be the case. However,
        it must be kept in mind, that the ejecta of 1994 may have cooled
        to  a lower temperature than 900K and hence  its peak emission will 
        not be in the  $K$ band but rather  at longer wavelengths.
        It may be pointed out, that the distance estimate of 300pc by 
       Martini et al (1994), based on the observed spectral class at outburst
       and derived intrinsic luminosity,  may be uncertain to some extent.
       Thus, a change in $d$  will affect the  derived dust-shell luminosity
        and mass (derived subsequently). However, the main
       conclusion that the  900K dust shell is not due to the 1994 
       outburst, should still remain valid. This conclusion -  based  on 
       comparison of the derived $R$${_{\rm shell}}$ from kinematic and 
       luminosity  arguments - should not be affected by any reasonable error
       in the  distance estimate.

               The mass of the dust shell can be calculated as $M$${_{\rm
        dust}}$ =1.1$\times$$10$${^{\rm 6}}$ $d$${^{\rm
        2}}$($\lambda$$F$${_{\rm \lambda}}$$)$${_{\rm max}}$ / $T$${_{\rm
        dust}}{^{\rm 6}}$ (Woodward et al. 1993). Here $M$${_{\rm dust}}$
        is in units of $M$$_\odot$, $T{_{\rm dust}}$ is in units of
        $10{^{\rm 3}}$K, $d$ is in kpc and $(\lambda
        F{_{\rm\lambda}}){_{\rm max}}$ is measured at the peak of the SED of
        the dust in Wcm${^{\rm -2}}$.  The dust is assumed to comprise
        of carbon particles having a size and density of
        $\le$1${\rm{\mu}}$m and 2.25 gm/cm${^{\rm 3}}$ respectively. By
        extrapolating the 900K graph of Figure 2 (lower panel) to longer
        wavelengths to determine $(\lambda F{_{\rm \lambda}}){_{\rm
        max}}$, we find $M{_{\rm dust}} = 3.66 \times 10{^{\rm
        -12}}$$M${$_\odot$} for T = 900K and $d$ = 300pc. For a canonical
        value of 200 for the gas-to-dust ratio, the mass of the gas is
        $M{_{\rm gas}} = 7.3 \times 10{^{\rm -10}}$$M${$_\odot$}.
        The assumption that carbon particles make up the dust shell, may 
       not be entirely valid since silicates are also  expected in view of the
       oxygen-rich atmosphere inferred from the AlO bands. However, the 
       contribution from silicates to the SED is expected at longer wavelengths 
       (Mason et al., 1998). On the whole however, there may be a dust
       component due  to silicates, an additional component of  cooler 
       temperature dust which will not be seen in the near-IR $JHK$ colors 
       here and also more matter  outside the dust shell -in the form of a 
       molecular shell - as discussed shortly. Hence the dust mass derived
       here may be considered a lower limit.

	Following the same approach as that for the dust shell, we derive
       the luminosity and radius of the source at the 2MASS epoch to be
       $L$${_{\rm \star}}$ $\sim$ 0.3$L$$_\odot$  and $R$${_{\rm \star}}$
        $\sim$ 1.6$R$$_\odot$.  The $T$${_{\rm eff}}$ = 3250K is indicative of
       an early M type star. However $L$${_{\rm \star}}$ and 
       $R$${_{\rm \star}}$ deviate considerably from those of a  
       giant/supergiant and are closer to a main sequence star (for a M0V star,
       $L$ = 0.08$L$$_\odot$; $R$ = 0.6$R$$_\odot$ ). This
       is indicative that the star has possibly evolved to its pre-outburst
       state. It is puzzling that the  2MASS luminosity is  lower than that
       of the dust shell seen presently. While the discrepancy is not large,
       it is expected that $L$${_{\rm dust}}$ should be lower than 
       $L$${_{\rm \star}}$ in case the dust is just reradiating the energy
       from the central source.

	\subsection{Modelling of the AlO bands and Discussion}
	        
        It is difficult to locate the site of the AlO emission.  Since the
        AlO bands are seen in emission they are not photospheric.
        Presumably it is mixed with the dust in the dust-shell or located
        outside it. The source of excitation of the bands is likely to be
        due to fluorescence by near-IR photons emitted from the star or
        from the dust shell (or both).  We are attempting to model the AlO
        bands in details in a separate work. However, we present here a
        preliminary synthetic spectrum (Figure 1), computed for a
        vibrational and rotational temperature T(vib) = 3000K and T(rot)=
        300K respectively. This spectrum gives a reasonable reproduction 
        of the relative band strengths. The unrelated values of T(vib) and
        T(rot) indicate that the system is not in thermal equilibrium. The
        model calculations, done for the (1,0) and (2,0) bands, indicate a
        low rotational temperature in the range of 200-400K range is required 
        to reproduce the observed AlO bands. This can be seen from Figure 3 
        where the (2,0) band profiles are shown for different values of T(rot).

               Tsuji (1973) has calculated the molecular abundances of
        different species in physical conditions similar to cool stellar
        atmospheres. These calculations show that the production of AlO is
        highest in the temperature range 2000-2500K.  It then reduces
        with decreasing temperature 
        and by 1000K the abundance drops to $10$${^{\rm -7}}$ times
        its peak value.  The above considerations indicate that AlO is
        produced at a site conducive for its generation i.e. the cool
        photosphere of V4332 Sgr where a temperature of 2300K prevailed
        shortly after the outburst. The generated AlO is subsequently
        lost, possibly in a wind, and is presently seen along with the
        dust shell. We rule out the possibility of AlO being created in
        the dust shell itself which is at too low a temperature (900K) for
        efficient AlO production.  The same process of AlO production may
        also apply for V838 Mon where appropriate temperatures exist (Munari
        et al. 2002, Banerjee \& Ashok 2002a; Crause et al. 2003). However
        an important distinction is that the AlO bands in V838 Mon are
        seen in absorption implying they are photospheric. They may appear
        in emission at a later stage. Tsuji's results (1973) show that
        AlH and AlOH (among other Al bearing molecules) may also be expected.
	However available data in the literature shows the AlH fundamental
        band lies at 6${\rm{\mu}}$m  and the first  overtone at 3${\rm{\mu}}$m 
        i.e. outside the spectral coverage reported here (White et al., 1993;
        Deutsch et. al, 1987).	For the AlOH isomer , there is only a theoretical
        paper (Hirota et al. 1993) where the O-H frequency is expected
	near 2.3${\rm{\mu}}$m. A laboratory confirmation of this should possibly
	be awaited before looking for such bands in the present data.  
          
               Although the outburst mechanism for V4332 Sgr/V838 Mon type
        of objects is not completely understood, available evidence 
        indicates that the eruption is basically a dramatic expansion 
        of the star into a supergiant accompanied by some mass loss in a
        moderate velocity wind (Bond et al. 2003). Here we have shown that
        the mass loss in such objects may continue beyond the main eruption.
        Based on the present results   on V4332 Sgr, it appears that 
        post-outburst developments in these objects can be fairly rapid
        and interesting (e.g. the rich AlO spectra) and thus there is a
        need to monitor them regularly. A cold
        molecular environment can surround these stars in their
        post-outburst stage in sharp contrast to the extremely hot coronal
        gas in classical novae.  Specifically in V4332 Sgr - given the
        predominace of AlO in the spectra - it would be worthwhile to look
        for the B-X band system of AlO in the optical and also for
        rotational transitions of the lower $X$$^{\rm 2}$$\Sigma$$^{\rm
        +}$ levels in the sub-mm/mm region where several such lines are
        listed.

	\acknowledgements
        The research work at PRL is funded by the Department of Space,
        Government of India.  We thank the UKIRT service program for
        observation time in the service mode and Chris J. Davis and Jane
        Buckle of UKIRT for doing the observations.  UKIRT is operated by
        JAC, Hawaii, USA, on behalf of the UK Particle Physics and
        Astronomy Research Council.

\clearpage


\begin{figure}
\plotone{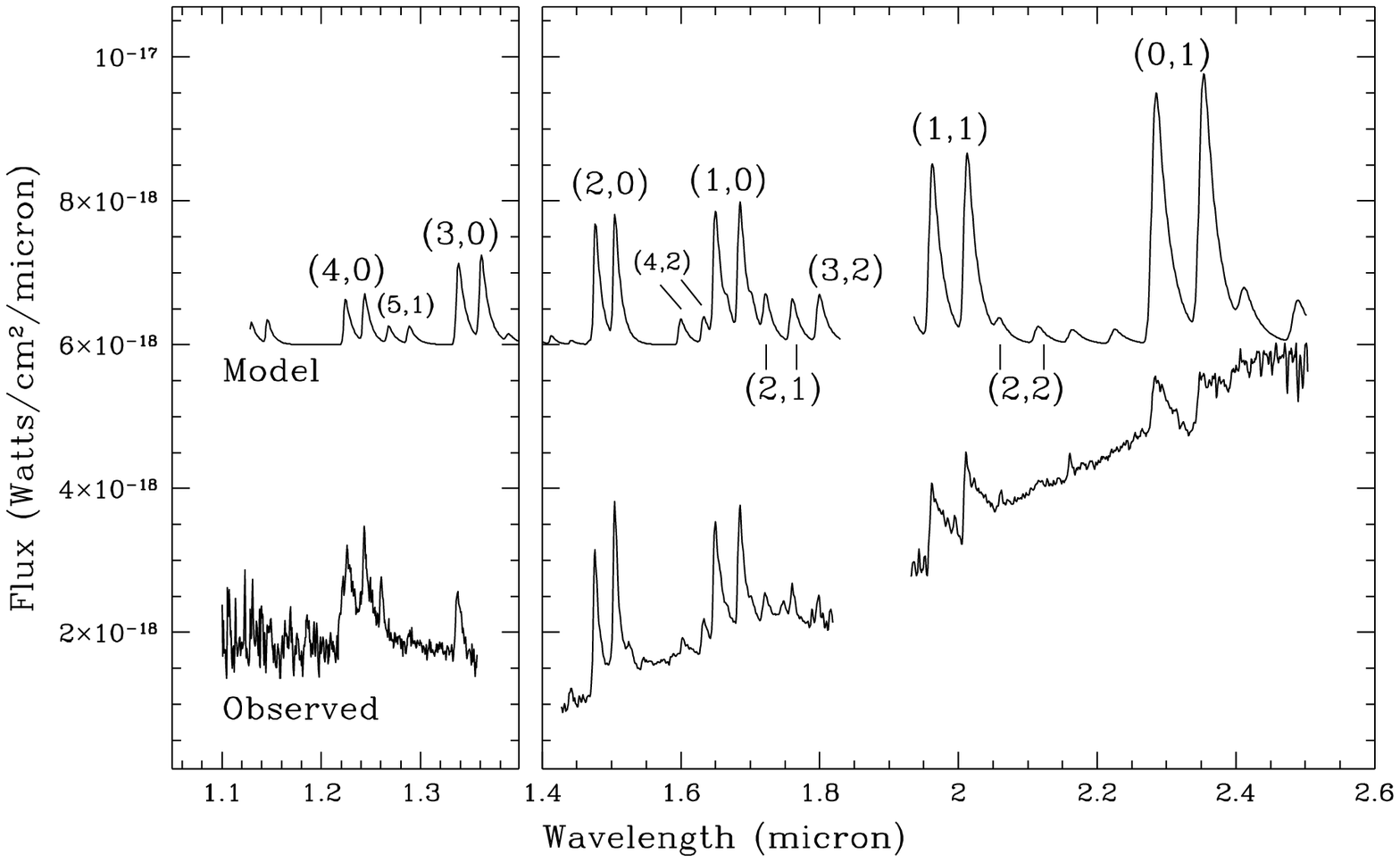}
\caption{The observed $JHK$ spectra of V4332 Sgr showing the prominent
$A$$^{\rm 2}$$\Pi$$_{i}$$-$$X$$^{\rm 2}$$\Sigma$$^{\rm +}$ emission bands 
of AlO are displayed in the bottom section. A computed, model spectrum is also
shown in the upper half for comparison. The identifications of the detected
bands are marked in the model spectrum. \label{fig1}}
\end{figure}

\clearpage 

\begin{figure}
\plotone{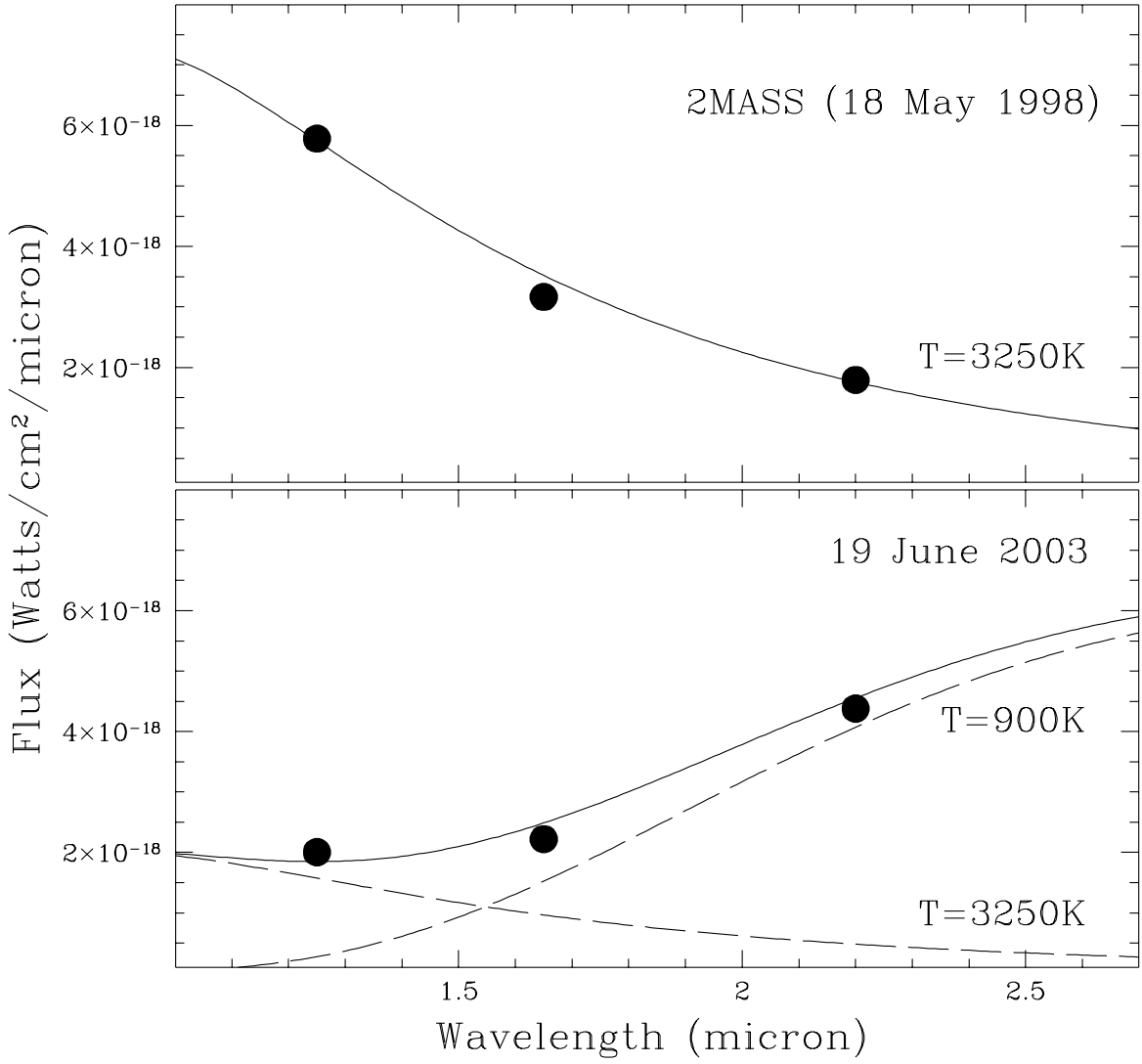}
\caption{The top panel shows a T = 3250K black-body fit to the 2MASS data.
The bottom panel, shows  the sum (bold line) of 2 black-body componemts 
at T = 900K and 3250K respectively (dashed lines) fitted to the present $JHK$
fluxes. \label{fig2}}
\end{figure}

\clearpage
\begin{figure}
\plotone{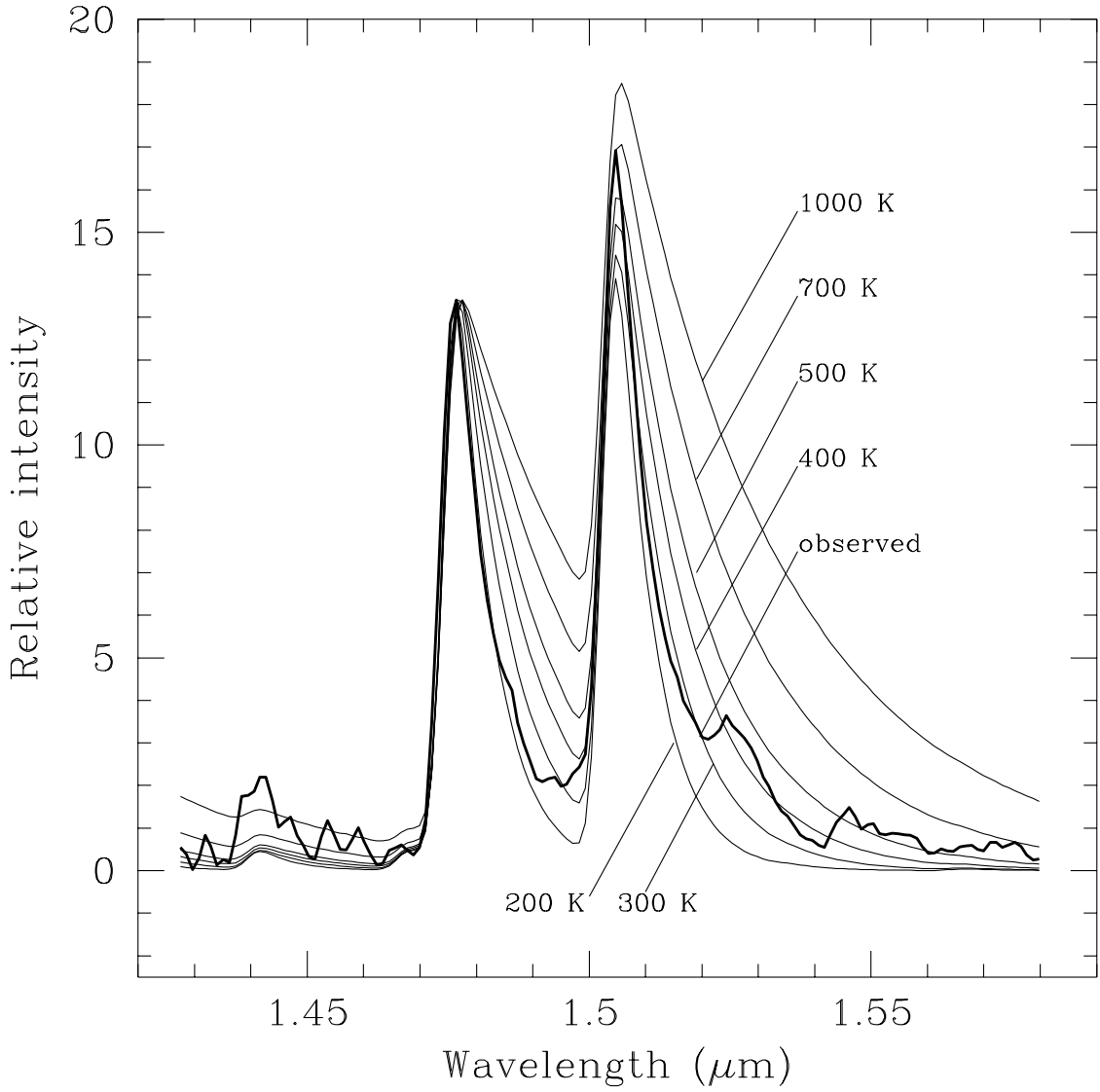}
\caption[]{ Model fits at different rotational temperatures for the (2,0) A-X
 band of AlO are shown.  Please refer the text for further 
 details. \label{fig3}}
\end{figure}





\clearpage
\begin{table}
\caption{Log of spectroscopic observations}
\begin{tabular}{lllllll}
\hline \\ 
UT Date   & UT & Grism & Resolution & Integration\\
          &    &       &            & Time(s)\\
\hline 
\hline 
24 April 2004 & 13.879   & HK      & 400-500  & 720  \\
24 April 2004 & 14.319   & Long J  &  2050    & 960  \\
4 May 2003    & 14.902   & IJ      &   300    & 1800 \\
\hline
\end{tabular} 
\end{table}

\clearpage  
\begin{table}
\caption{Log of photometric observations for 19 June 2003}
\begin{tabular}{lllllll}
\hline \\ 
UT    & Band & Exposure & Integration & Mag. (error)\\
          &      & Time(s)  & Time(s)     &             \\
\hline 
\hline 
12.199 & J   & 50  & 250  & 13.251 (0.005) \\
12.017 & H   & 10  & 90   & 11.986 (0.01)  \\
11.955 & K   & 4   & 80   & 10.023 (0.011) \\
\hline
\end{tabular} 
\end{table}

\clearpage
\begin{table}
\caption{A list of the laboratory wavelengths for the
 $A$$^{\rm 2}$$\Pi$$_{i}$$-$$X$$^{\rm 2}$$\Sigma$$^{\rm +}$ transitions of AlO}
\begin{tabular}{lllllll}
\hline \\ 
Band   &$\lambda$(${\rm{\mu}}$m)& $\lambda$(${\rm{\mu}}$m) & Band & 
$\lambda$(${\rm{\mu}}$m) & $\lambda$(${\rm{\mu}}$m)       \\
 & $A$$^{\rm 2}$$\Pi$$_{\rm 3/2}$  & $A$$^{\rm 2}$$\Pi$$_{\rm 1/2}$  & & $A$$^{\rm 2}$$\Pi$$_{\rm 3/2}$  & $A$$^{\rm 2}$$\Pi$$_{\rm 1/2}$ \\
\hline 
\hline \\ 

(1,0) & 1.6837 & 1.6480 & (1,1) & 2.0106 & 1.9600 \\
(2,0) & 1.5035 & 1.4749 & (2,1) & 1.7589 & 1.7199 \\
(3,0) & 1.3589 & 1.3361 & (5,1) & 1.2874 & 1.2669 \\
(4,0) & 1.2425 & 1.2256 & (3,2) & 1.8409 & 1.7962 \\
(0,1) & 2.3514 & 2.2826 & (4,2) & 1.6311 & 1.6031 \\
 
\hline
\end{tabular} 
\end{table}


\begin{thebibliography}{}

    
\bibitem[ref1]{r1} Banerjee, D.P.K., $\&$  Ashok, N.M.,
2002a, A$\&$A, 395, 161   


\bibitem[ref2]{r2}
Banerjee, D.P.K., $\&$ Ashok, N.M., 
2002b, IAU Circ. 8036 

\bibitem[ref19]{r19}  
Barnbaum, C., Omont, A., $\&$ Morris, M.,
1996, A$\&$A, 310, 250


\bibitem[ref3]{r3}
Bernstein, L.S., Rudy, R.J., Lynch, D.K., Dimfl, W.L., Mazuk, S.,
Venturini, C.C., Puetter, R.C., $\&$ Perry, R.B., 2003, IAU Circ. 8082 

\bibitem[ref4]{r4}
Bond, H.E., Henden, A., Levay, Z.G., Panagia, N., Sparks, W.B., Starrfield, S.,
Wagner, R.M., Corradi, R.L.M., $\&$ Munari, U., 2003, Nature, 422, 405

\bibitem[ref5]{r5}
Crause, L.A., Lawson, W.A., Warrick, A., Kilkenny, D., Van Wyk, F.,
Marang, F., $\&$ Jones, A.F.,  2003, \mnras, 341, 785


\bibitem[ref16]{r16}
Deutsch, J.L., Neil, W.S., $\&$ Ramsay, D.A.,
  1987, J. Mol. Spectrosc., 125, 115
   
\bibitem[ref6]{r6}
Geballe, T.R., Smalley, B., Evans, A. $\&$ Rushton M.T.,
  2002, IAU Circ. 8016


\bibitem[ref15]{r15}
Hirota, F., Tanimoto, M., $\&$ Tokiwa, H.,
  1993, Chem. Phys. Letters, 208, 115


\bibitem[ref7]{r7}
Koornneef, J., 1983, A$\&$A, 128, 84

\bibitem[ref8]{r8}
Launila, O., $\&$ Jonsson, J.,
  1994, J. Mol. Spectrosc., 168, 1 

\bibitem[ref9]{r9}
Martini, P., Wagner, R.M., Tomaney, A., Rich, R.M., Della Valle, M.,
$\&$ Hauschildt, P.H., 1999, \aj, 118, 1034


\bibitem[ref18]{r18}
Mason, C.G., Gehrz, R.D., Woodward, C.E., Smilowitz, J.B., Hayward, T.L.,
 $\&$ Houck, J.R.,
  1998, \apj, 494, 783


\bibitem[ref10]{r10}
Munari, U., Henden, A., Kiyota, S., Laney, D., Marang, F., Zwitter, T.,
Corradi, R.L.M., Desidera, S., Marrese, P., Giro, E., Boschi, F., $\&$
Schwartz, M.B.,  2002, A$\&$A, 389L, 51 


\bibitem[ref11]{r11}
Rich, R.M., Mould, J., Picard, A., Frogel, J.A., $\&$ Davies, R.,
  1989, \apj, 341, L51

\bibitem[ref12]{r12}
Soker, N., $\&$ Tylenda, R., 
  2003, \apj, 582, L105


\bibitem[ref13]{r13}
Tsuji, T.,
  1973, A$\&$A, 23, 411
  
\bibitem[ref17]{r17}  
White, J.B., Dulick, M., $\&$   Bernath, P.F.,
1993, J. Chem. Phys., 99, 8371

\bibitem[ref14]{r14}
Woodward, C.E., Lawrence, G.F., Gehrz, R.D., Jones, T.J., Kobulnicky, H.A., 
Cole, J., Hodge, T., $\&$ Thronson (Jr.), H.A.,  1993, \apj, 408, L37



\end{thebibliography}
\end{document}